\documentclass[conference]{IEEEtran}
\IEEEoverridecommandlockouts
\usepackage{cite}
\usepackage{amsmath,amssymb,amsfonts}
\usepackage{algorithmic}
\usepackage{graphicx}
\usepackage{textcomp}
\usepackage{xcolor}
\def\BibTeX{{\rm B\kern-.05em{\sc i\kern-.025em b}\kern-.08em
    T\kern-.1667em\lower.7ex\hbox{E}\kern-.125emX}}
\begin{document}

\title{Editable AI: Mixed Human-AI Authoring of Code Patterns \\
%{\footnotesize Anonymized for Submission} \\
}

\author{\IEEEauthorblockN{Kartik Chugh}
\IEEEauthorblockA{\textit{Department of Computer Science} \\
\textit{University of Virginia}\\
Charlottesville, VA, USA \\
kc6afx@virginia.edu}
\and
\IEEEauthorblockN{Andrea Y. Solis}
\IEEEauthorblockA{\textit{Department of Computer Science} \\
\textit{George Mason University}\\
Fairfax, VA, USA \\
asolis6@gmu.edu}
\and
\IEEEauthorblockN{Thomas D. LaToza}
\IEEEauthorblockA{\textit{Department of Computer Science} \\
\textit{George Mason University}\\
Fairfax, VA, USA \\
tlatoza@gmu.edu}
}

\IEEEpubid{\makebox[\columnwidth]{978-1-7281-0810-0/19/\$31.00~\copyright2019 IEEE \hfill} \hspace{\columnsep}\makebox[\columnwidth]{ }}

\maketitle

\IEEEpubidadjcol

\begin{figure*}[t!]
% \centerline{}
\centering
\includegraphics[width=2.0\columnwidth]{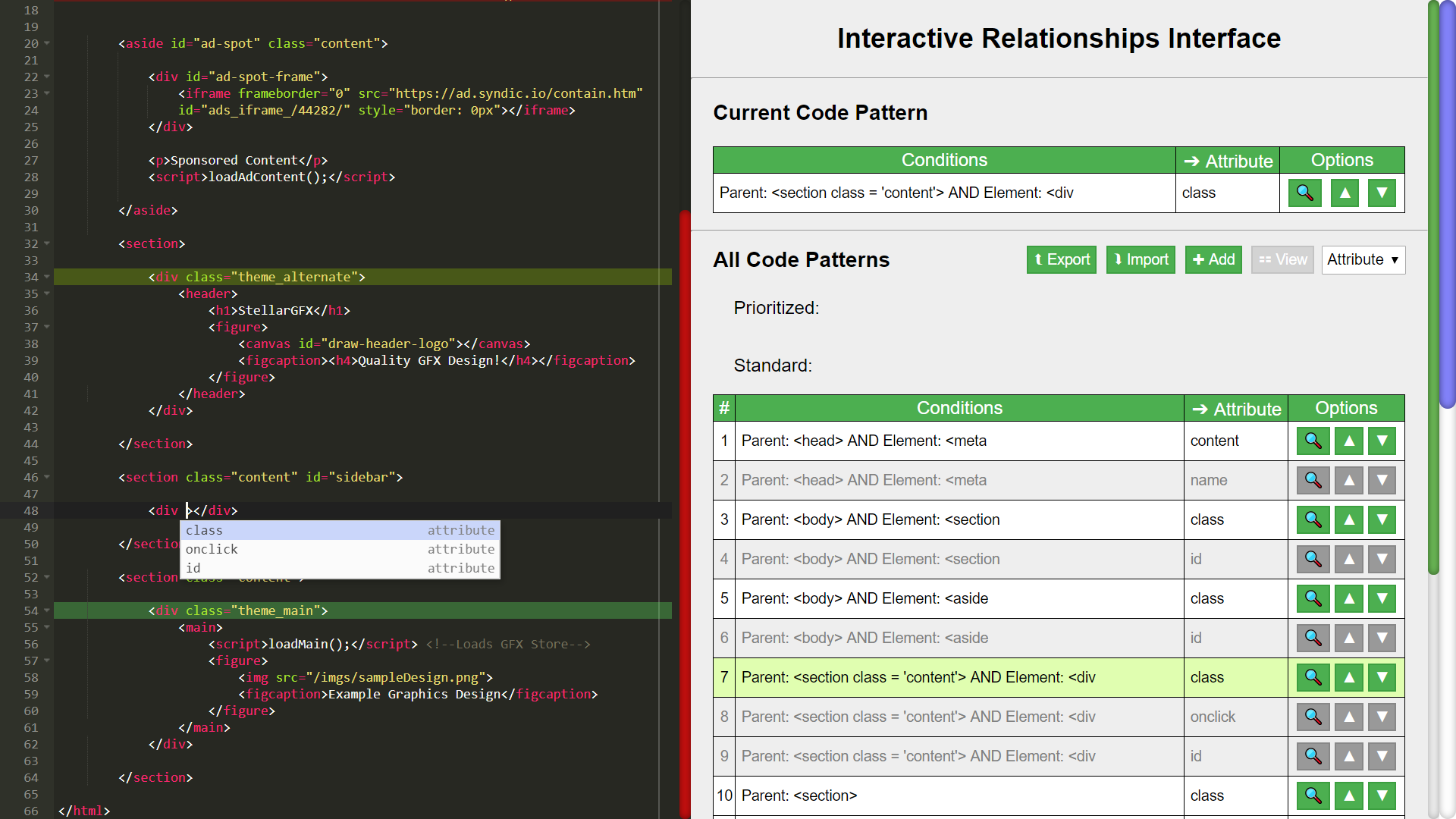}
\caption{As developers type in the HTML editor (left side), the autocomplete menu is displayed to offer suggested code completions. In the IRIS interface (right side), the Current Code Pattern explains to the developer the basis for the first suggested completion.}
\label{fig:alice_div}
\end{figure*}

\begin{abstract}
Developers authoring HTML documents define elements following patterns which establish and reflect the visual structure of a document, such as making all images in a footer the same height by applying a class to each. To surface these patterns to developers and support developers in authoring consistent with these patterns, we propose a mixed human-AI technique for creating code patterns. Patterns are first learned from individual HTML documents through a decision tree, generating a representation which developers may view and edit. Code patterns are used to offer developers autocomplete suggestions, list examples, and flag violations. To evaluate our technique, we conducted a user study in which 24 participants wrote, edited, and corrected HTML documents. We found that our technique enabled developers to edit and correct documents more quickly and create, edit, and correct documents more successfully. 
\end{abstract}


\section{Introduction}
Developers authoring HTML documents define elements in ways which reflect patterns. For example, a developer might describe a navigation control by adding a number of button elements as children of a \texttt{<div>} tag, ensuring each button has a similar class which establishes its visual style and enables its association with logic in code. While elements in documents may be styled through Cascading Style Sheets (CSS), which describe visual properties which can be applied to elements, HTML documents exhibit patterns which reflect their structure. For example, a \texttt{div} element might contain only \texttt{link} elements, each with the same attribute. As developers work with HTML documents, developers often wish to make edits consistent with the existing structure. Developers might be supported in this activity through autocomplete suggestions, suggesting elements or attributes which reflect the document's structure, or might be informed when they write documents inconsistent with this structure. Offering this support requires a model of the code patterns which exist in the document.

Machine learning systems offer the possibility of identifying patterns from data. For example, a model trained on an HTML document might suggest that the most likely child element for a \texttt{div} container with a \texttt{class} value of \texttt{nav-bar} is an \texttt{img}. However, traditional machine learning systems lack explainability, offering no ability for the user to understand why the prediction was made. Moreover, in contexts where the patterns to be learned from data reflect patterns that the user themselves intended to create, the user may have a better model of expected behavior than the data itself. But traditional machine learning approaches lack editability, making it impossible for the user to correct or edit learned patterns to reflect their intent.

%Machine learning systems identify patterns in data, supporting users by offering predictions and identifying potential outliers. By finding and applying patterns in ways that humans cannot, systems bring new insights which benefit users, powering recommendation systems, forecasting tools, and... [other systems.]

%Yet implicit in this interaction is that the data knows best: whichever model best fits the data is the model to use. But there may be situations in which users have insight into the data beyond the data itself. When the user themselves has authored the data, patterns may reflect the user's own intent, which the user may know better than the data. For example, consider a user editing a document who creates a list of items, ensuring that each item 

%Existing systems lose directness and immediacy, in that user must interact with a secondary representation rather than the document itself.

We envision a new form of human-computer collaboration in which the human (a developer) works together with the computer (the IDE) to author patterns reflecting a document's structure. A machine learning algorithm is first used to identify patterns. Using the model, the computer offers the developer code predictions, helping them complete tasks more quickly, and flags anomalies, helping them correct potential mistakes. When the code patterns learned from the data do not reflect the developer's true intent, the developer may view a representation of the computer's model, editing the model to reflect their intent. 

We explore this approach in the context of a developer editing HTML documents. Patterns reflecting the structure of HTML elements and attributes are learned from individual HTML documents using a decision tree. As developers create new elements in the document, autocomplete suggests tag names and attributes based on the model. Developers may view the underlying model, viewing individual patterns identified (e.g., a parent tag of \texttt{<head>} and element tag of \texttt{<meta>} implies a \texttt{content} attribute). Developers may then see examples of each pattern, find code snippets which violate the pattern, and edit the pattern to better reflect their intent. We implemented this approach in a prototype tool called IRIS (Interactive Relationships Interface System), an extension to an HTML editor which enables developers to interact directly with code patterns.

To evaluate our approach, we conducted a user study in which 24 participants wrote, edited, and corrected HTML documents. We found that IRIS enabled participants to edit and correct documents more quickly and create, edit, and correct documents more successfully. Developers used IRIS to understand the source of autocomplete suggestions, developing trust in the system, as well as to identify examples of code patterns to better understand them.

%AI systems can find patterns. 
%Humans can find patterns. 
%Want to use both of these together.

%Explainable AI.
%Modifiable AI.

%In this paper, we explore a technique for 

%[building on the most similar idea from existing work]
%Retaining the ability of a direct manipulation interaction in which a user can at any time edit any content, users

%Workflow of infer, suggest, review, promote.

%System infers rules from patterns.
%System users patterns to suggest new styling through autocomplete.
%User may review styling rules

%User retains full control over the inference of rule process.

%Finds outliers. Flexibile, in that 

\section{Motivating Example}

\begin{figure}[tp]
% \centerline{}
\centering
\includegraphics[width=1.0\columnwidth]{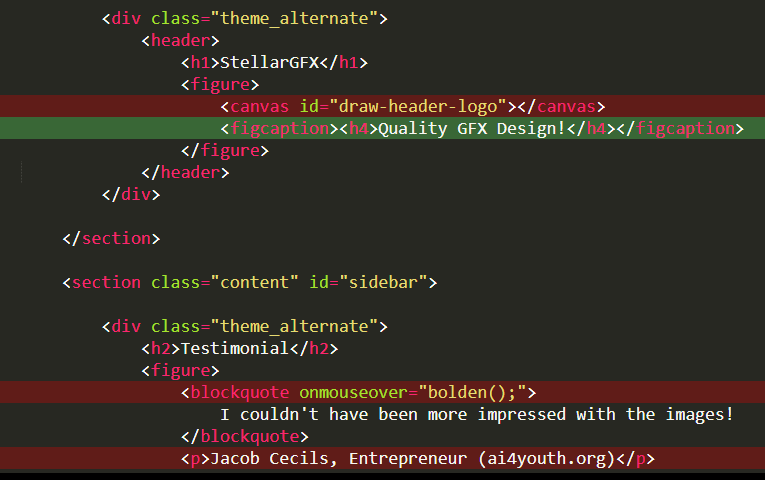}
\caption{Invoking the pattern inspector highlights examples of the code pattern (in green and yellow) and violations (in red) in the HTML editor. In this example, Alice is inspecting the pattern that \texttt{figure} parents contain \texttt{figcaption} children. Alice can see that the \texttt{figure} she is creating in the bottom 5 lines differs from the existing \texttt{figure} Bob created above, as hers uses a \texttt{<p>} tag in place of a \texttt{<figcaption>} tag.}
\label{fig:alice_figcaption_focused}
\end{figure}

\begin{figure}[tp]
% \centerline{}
\centering
\includegraphics[width=1.0\columnwidth]{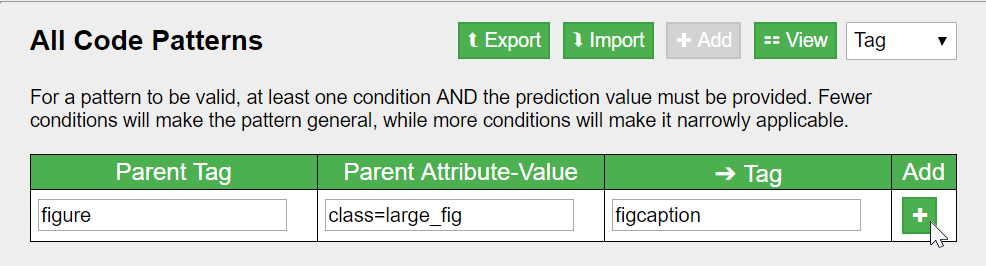}
\caption{Authoring a custom code pattern involves specifying a target feature and the condition feature(s) that imply it. In this example, Alice creates a code pattern that a \texttt{figure} with a \texttt{class} of \texttt{large\_fig} should have a \texttt{figcaption} child element.
}
\label{fig:alice_add}
\end{figure}

Alice recently joined the web development team of a graphics design company. She has been tasked with completing the company's new webpage, which has been partially developed by her co-worker Barry. Alice loads his HTML document into an IRIS-augmented editor and scrolls down to the live preview panel to assess the current progress on the webpage. Observing its abundant white space and empty look, she decides to create a sidebar on the left side of the page.

As Alice types code which styles the sidebar, an autocomplete menu suggests several HTML attributes to apply, ordered by the system's confidence in each suggestion (see \figurename{ \ref{fig:alice_div}}). Curious to understand why the first attribute was recommended to her, she looks at the Current Code Pattern panel, which describes the HTML features (``conditions'') that guided the top attribute recommendation. In this case, Alice learns that a \texttt{<div>} tag---nested under a \texttt{<section class="content">} parent element---suggests that a \texttt{class} attribute follows. Alice clicks the magnifying glass icon, highlighting examples of the code pattern in the HTML editor (an activity we name ``pattern inspection'' in this paper). Reading a few of these examples, Alice realizes that Barry usually applied a \texttt{class} attribute to elements like these, and sees examples of several of the classes he applied. With this insight, Alice mimics it in her sidebar, achieving a cohesive visual design.

After adding a client testimonial to the sidebar, Alice wants to verify that her new code follows the same general structure as Barry's. She selects Tag from the All Code Patterns dropdown and browses the list for patterns concerning related elements. For each such pattern, she uses the pattern inspection tool to see examples as well as violations in the HTML document. For example, in one code pattern Alice reads that a \texttt{figure} element contains a \texttt{figcaption} child element. The pattern inspector reveals that her \texttt{figure} instead uses a \texttt{p} element (highlighted in red), violating Barry's code pattern (see \figurename{ \ref{fig:alice_figcaption_focused}}). Alice fixes her mistake by copying one of the \texttt{figcaption} pattern's usage examples (highlighted in green) and tweaking the caption to fit her image. 

Having completed the sidebar, Alice intends to implement a footer displaying the logos of the company's partners. Alice decides that the \texttt{figcaption} pattern is overly broad and does not want it to apply to the logo images. To narrow its applicability, Alice downvotes the pattern, removing it from consideration, and clicks the Add button to write a more specific version: \texttt{<figcaption>} is the child tag of a \texttt{<figure class="large\_fig">} parent element (see \figurename{ \ref{fig:alice_add}}). She then adds this attribute-value pair to the captioned \texttt{<figure>} elements already in the document to comply with her new pattern. The system no longer recommends that Alice include \texttt{figcaption} inside unclassed \texttt{figure}s, and will not flag her footer code for failing to do so. In this way, Fred, a future developer working in the same document to continue Alice's work, can be made aware of Alice's caption pattern. 

Alice may also choose to share her code patterns with Michele. Michele is working on a different document, but wishes to use a similar look and feel. Alice can first use the Export button to download a JSON file containing the document's code patterns and send these to Michele. Michele can then click Import to import these code patterns into her document.

\section{System} 
In IRIS, the computer and the human work together to create code patterns. The computer first learns code patterns from the HTML document by training decision trees. These decision trees are then used to offer the developer potential completions using an autocomplete interface, suggesting potential tags, attributes, or values. Developers may then interact directly with the code patterns, using a dedicated interface to view the code pattern responsible for a specific recommended completion, edit code patterns, and browse examples of code patterns. IRIS is implemented as an extension to a simple web-based HTML editor. In the following sections, we describe how IRIS learns code patterns from HTML documents, how code patterns are used to suggest code completions, and how developers may interact directly with code patterns.
%IRIS is a browser based HTML editor augmented with support for working with code patterns. In this section, we define the "system" to include both the visible user interface as well as the JavaScript program powering its functionality. IRIS consists of an editor panel on the left, and the IRIS interface on the right, which is divided into the "Current Code Pattern" and "All Code Patterns" panels. (\figurename{\ref{fig:alice_div}}).

\subsection{Learning Code Patterns}

\begin{figure}[tp]
% \centerline{}
\centering
\includegraphics[width=0.7\columnwidth]{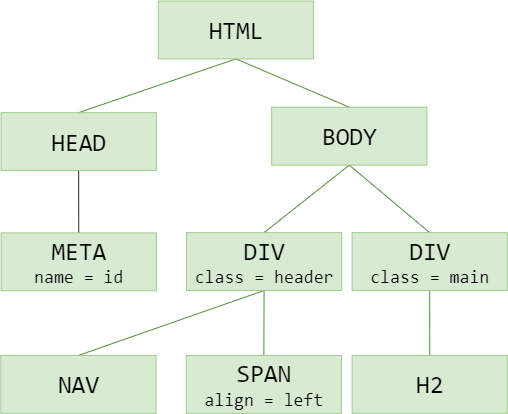}
\caption{IRIS parses HTML documents into an AST, where each node corresponds to a tag and may be associated with zero or more attributes and attribute values.}
\label{fig:model_ast}
\end{figure}

In this paper, we focus on HTML documents as individual tokens of HTML. \textit{Tags}, \textit{attributes}, and attribute values, or just \textit{values}, are three important HTML tokens. Code patterns represent a relationship between between \textit{condition features} and \textit{target features}, where the former implies the latter. 
%Because our system learns patterns tags, attributes, and attribute values, they also represent the target feature types. Our system also uses \textit{parent tags} and \textit{parent attribute-values} as additional context features.  

To learn code patterns, IRIS first builds an abstract syntax tree (AST) for the active HTML document using Himalaya.js\footnote{https://github.com/andrejewski/himalaya}(shown in \figurename{ \ref{fig:model_ast}}). 
%Each node in the AST corresponds to an HTML element, and HTML elements may have zero or more attributes or attribute-value pairs. 
As code patterns may or may not be document-specific, by default all code patterns are learned from and applied to an individual document. 

\begin{figure}[tp]
% \centerline{}
\centering
\includegraphics[width=0.8\columnwidth]{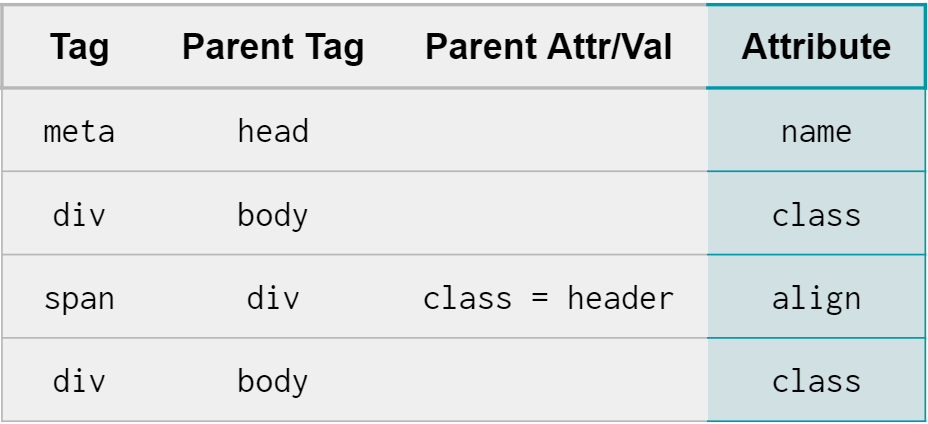}
\caption{IRIS constructs training data tables for each of the decision trees for tags, attributes, and attribute values. 
The training data for attributes, shown here, includes the enclosing tag, parent tag, and its attribute-value pairs, as well as the resulting target attribute.}
\label{fig:model_training}
\end{figure}

IRIS constructs three separate decision trees corresponding to the three target types it may predict: an HTML tag, attribute, or value. 
IRIS bases its predictions on \textit{features} extracted from the AST. In examining code patterns, we found that many patterns are contextual, reflecting the role of an element within the HTML document. For example, a tag contained in a \texttt{div} with a \texttt{class="sidebar"} attribute-value pair might vary from a tag contained in a \texttt{table}. Thus, the key condition features we chose to extract are the parent element tag and its attribute-value pairs. Additionally, for attribute and value targets, the features considered include the tag of the element that the developer is currently completing. When predicting values, the preceding attribute name of the current element is considered. IRIS collects the relevant features from the AST, populating a table of training data (shown in \figurename{ \ref{fig:model_training}}) for the target type. 
After gathering the training data, IRIS constructs a decision tree for each target type. Using a JavaScript implementation\footnote{https://github.com/willkurt/ID3-Decision-Tree} of the ID3 algorithm\cite{Quinlan1986}, a decision tree is learned from the training data. 

%In order to learn high-level code patterns from the document, the system must first aggregate low-level relations between pieces of code; these relations associate code conditions to targets exactly as they appear in the document. Recall from \ref{subsubsect:tokenization} that the type of target feature the system collects relations for is determined through tokenization. Furthermore, the condition feature types sought out by the system are dependent on the target feature type: consider the fact that tags can help predict a yet-to-be-typed attribute, but not vice versa. \textbf{[Insert table/image showing which condition types are relevant for each target type?]}

%Given the identified target type, the system recurses through the AST to extract the appropriate target features and the conditions associated with them (\figurename{\ref{fig:model_ast}}). These relations between conditions and targets, derived from within and among the AST nodes, form the training set for decision tree learning.

\begin{figure}[tp]
% \centerline{}
\centering
\includegraphics[width=1.0\columnwidth]{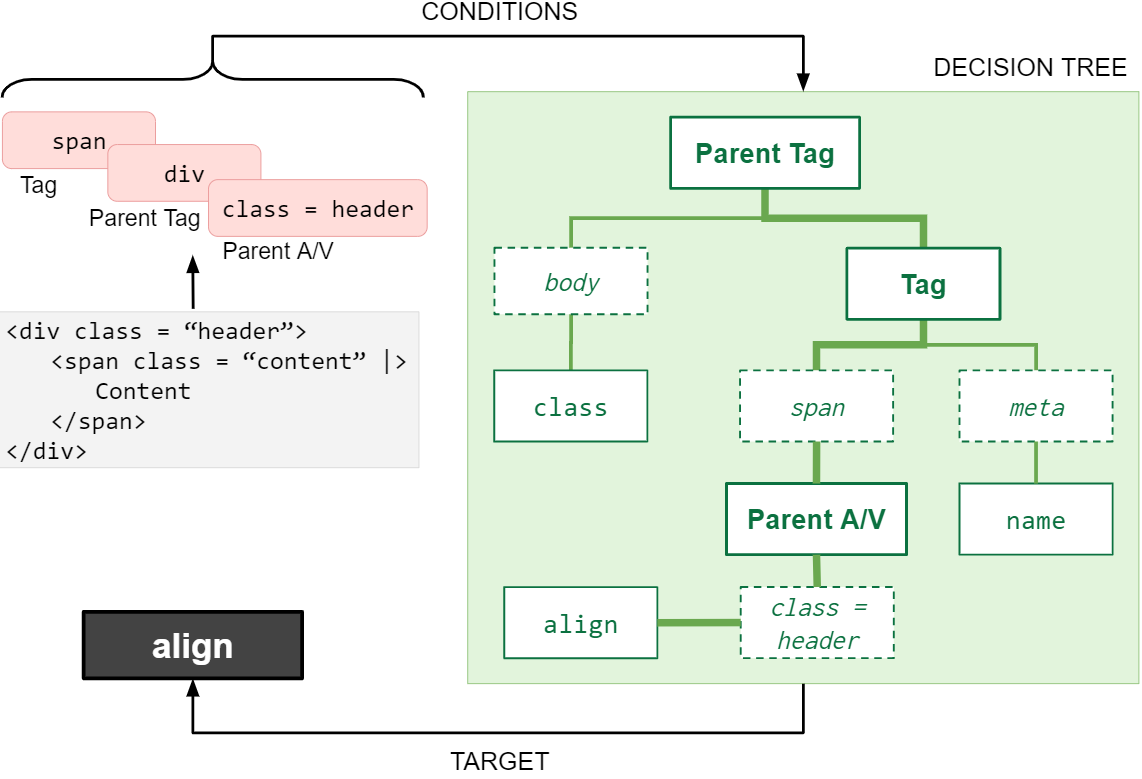}
\caption{Condition features are extracted from the current code context and used by the decision tree to predict targets. Only one target per decision node is shown for simplicity. The tree path taken in this example is bolded.}
\label{fig:model_dt}
\end{figure}

%The DT synthesizes the document's individual relations into a set of more generally-applicable "if-then" patterns, effectively mapping every possible combination of code conditions to targets. 
%ID3 was modified to allow for multiple targets ranked by probability of corresponding .

\subsection{Using Code Patterns to Suggest Completions}
\label{subsubsect:autocomplete}

As the developer enters each character in the HTML editor, IRIS offers the developer autocomplete suggestions. To determine the type of target (tag, attribute, or attribute value) to be completed, IRIS first tokenizes the characters in the current line, from the first character to the cursor position.   
%Code, spanning from the beginning of the current line being typed on to the user's cursor position, is grouped and classified into atomic units of code, or "tokens". 
IRIS then uses the two preceding tokens to determine the type of the target. Consider the following example (\texttt{|} represents the cursor position and \texttt{whsp} indicates whitespace):

\begin{quote}
    \texttt{<span class="content" |>}
\end{quote}

is tokenized as:

\begin{quote}
    \texttt{(tag)(whsp)(attribute)(value)(whsp)
    }
\end{quote}

\noindent
IRIS determines that a value followed by whitespace indicates that the next token may be an additional attribute. The target type is thus attribute. %Similar rules are applied to identify the type as tag, attribute, or value.

%\begin{figure}[htbp]
% \centerline{}
%\centering
%\includegraphics[width=1.0\columnwidth]{figures/iris_auto.png}
%\caption{Code patterns are  multiple decision tree-provided code completions.}
%\label{fig:iris_auto}
%\end{figure}

After determining the target type, IRIS consults the decision tree for this target type to generate potential completions. To ensure that the decision tree reflects the current version of the code, IRIS lazily trains the appropriate decision tree on demand. IRIS then retrieves the information about the context of the current cursor position. This includes the parent tag, attributes, and attribute values as well as the current tag and attribute, if applicable. With this data, IRIS then uses the decision tree to generate possible completions (\figurename{ \ref{fig:model_dt}}). These completions are displayed to the developer through an autocomplete interface (\figurename{ \ref{fig:alice_div}}). 
%The most immediate use of the decision tree is for predictive code completion. The system feeds the tree ``current code conditions'' to determine corresponding targets. These condition features are obtained from the HTML element the user is working on, as well as its parent element. The current conditions are entered as inputs to the decision tree, generating one or more code targets (\figurename{\ref{fig:model_dt}}). 
%To this point, the system has identified the type of HTML feature the user intends to code, intuited relationships between this and other feature types, and determined the target features implied by relevant relationships; it now displays these target features as autocomplete recommendations in a small dropdown menu 
%(shown in \figurename{\ref{fig:iris_auto}}).
%The user can use keyboard shortcuts or the mouse cursor to quickly perform code completion using any of the recommendations.

\subsection{Interacting with Code Patterns}

\begin{figure}[tp]
% \centerline{}
\centering
\includegraphics[width=1.0\columnwidth]{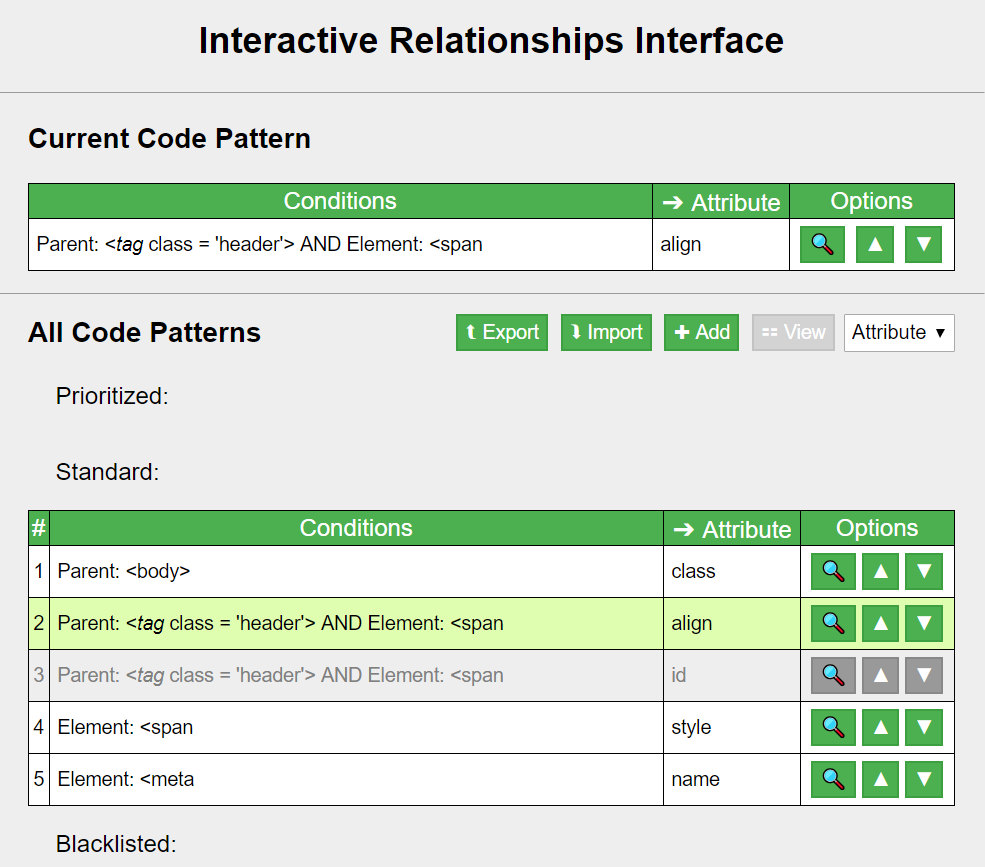}
\caption{The IRIS interface displays the code pattern for the current first listed autocomplete suggestion (if applicable) and tables of all code patterns.}
\label{fig:iris_patterns_panel}
\end{figure}

\subsubsection{Viewing Code Patterns}

To help developers understand code patterns and explain code completions, developers may directly view the set of code patterns. To offer developers a compact and understandable representation, we chose to present code patterns to the developer as a table of IF/THEN rules. Each rule describes a set of conditions in which it applies and the resulting predicted target. From each decision tree, each path through the decision tree, representing a series of conditions and a predicted target, is represented as a distinct code pattern.  
%In addition to real-time code completion, the decision tree is used for pattern extraction. As the system traverses the decision tree, it tracks the path it has taken. When a decision node is reached, the system constructs a pattern relating the path (in essence, a set of condition features) to the target feature. 
If multiple targets exist in a decision node, code patterns are constructed with identical conditions and the respective target. These patterns are stored in the target order expressed in the decision node, so as to preserve the ranking of more prevalent targets above less prevalent ones. 
%Together, all patterns extracted from every possible path of the decision tree form the standard patterns list.

Code patterns are displayed to the developer in the IRIS interface to the right of the HTML editor (\figurename{ \ref{fig:iris_patterns_panel}}). When autocomplete is active, the Current Code Pattern at the top describes the code pattern associated with the top autocomplete recommendation.
Below, tables for All Code Patterns display the code patterns for the document to the developer.
The developer may toggle between viewing code patterns for tags, attributes, and attribute value targets using the dropdown in the upper-right of the All Code Patterns section. 
Each pattern entry lists the conditions under which the code pattern applies and the predicted target tag, attribute, or attribute value. 
The current code pattern is indicated in the All Code Patterns section with a light green background. 

To enable developers to better understand the meaning and use of each code pattern, IRIS enables developers to see examples of each code pattern in the HTML editor. Clicking the magnifying glass next to the rule invokes the pattern inspector, highlighting examples of the code pattern. 
%understand the use and context of code patterns, each pattern entry includes a magnifying glass. Invoking the magnifying glass highlights usage instances, i.e., portions of code where the pattern manifests. 
If the example is a positive example, where the conditions and prediction match exactly, it is highlighted in green. If the conditions are similar but unequal, the example is highlighted in yellow. Conditions are considered similar if they differ only in one or more of the parent element's attribute-value pairs. Developers are also shown examples which violate the rule, shown in red, indicating a potential defect in document structure, visual style, or both. 

\subsubsection{Editing Code Patterns}

\begin{figure}[tp]
% \centerline{}
\centering
\includegraphics[width=1.0\columnwidth]{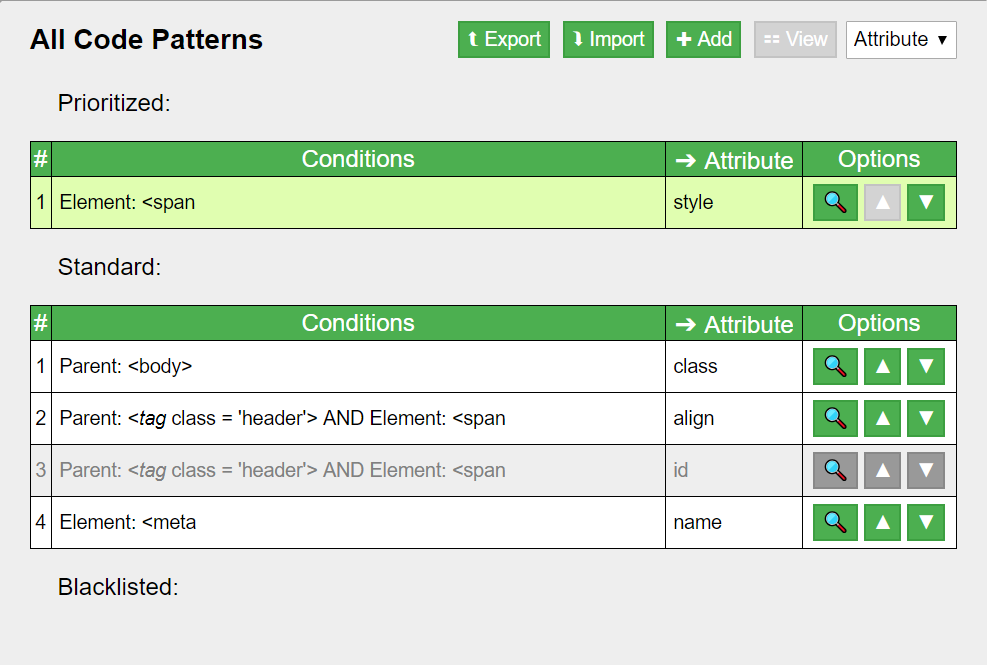}
\caption{Promoting a standard pattern moves it to the list of prioritized patterns.}
\label{fig:iris_patterns_priority}
\end{figure}

\begin{figure}[tp]
% \centerline{}
\centering
\includegraphics[width=1.0\columnwidth]{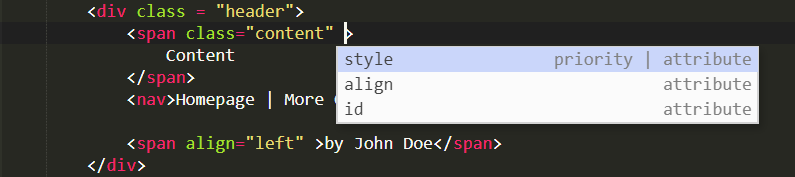}
\caption{Prioritizing patterns elevates their target features to the top of the autocomplete menu.}
\label{fig:iris_auto_priority}
\end{figure}

\begin{figure}[tp]
% \centerline{}
\centering
\includegraphics[width=1.0\columnwidth]{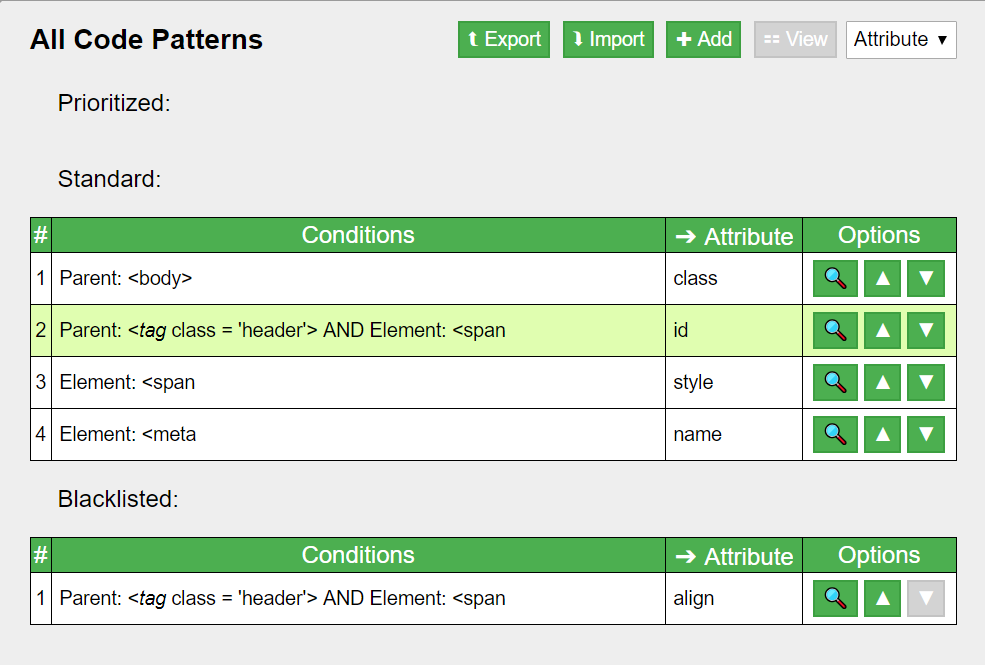}
\caption{Demoting a standard pattern moves it to the blacklist.}
\label{fig:iris_patterns_blacklist}
\end{figure}

\begin{figure}[tp]
% \centerline{}
\centering
\includegraphics[width=1.0\columnwidth]{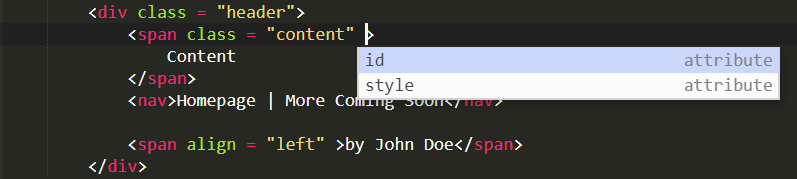}
\caption{Blacklisting patterns hides their target features from the autocomplete menu even if the pattern is applicable in the current context.}
\label{fig:iris_auto_blacklist}
\end{figure}

When inspecting code patterns, developers may identify patterns that they believe to be correct or incorrect or wish to manually create new code patterns expressing their intent. In these cases, the developer has insight into code patterns which they may offer to the computer. By taking the time to express their intent, developers may then receive suggested completions that better reflect the patterns they expect the document to follow.

In some cases there may be conflicting code patterns. Conflicting code patterns occur when there are multiple code patterns with identical conditions but which make different predictions. This reflects a HTML document which is inconsistent. In these cases, IRIS groups these code patterns together. The code pattern that is most likely is listed first and alternative code patterns are listed next and presented with a gray background and faded font. By grouping related code patterns, the developer may see alternatives and choose to indicate which they believe best reflects their intent. 

%which, given the same conditions, make Entries with the top target feature for a given set of conditions are emphasized, while entries with lower-ranked targets are presented in a faded font. This surfaces the patterns the user is most likely to interact with (e.g., reprioritizing a pattern for a top autocomplete recommendation). 

Developers may indicate that they believe a code pattern to be correct by upvoting the pattern (up arrow icon) or to be incorrect by downvoting the pattern (down arrow icon). This may be indicated from both the current code pattern and the list of All Code Patterns. Upvoting or downvoting a code pattern toggles it between three states: standard, prioritized, and blacklisted. Developers may assign any priority to any code pattern at any time. 

%From both the Current Code Pattern and All Code Patterns panels, users can promote or demote code patterns, which adjusts their priority levels within the system. Patterns may be in one of three priority levels: standard, prioritized, and blacklisted. Pattern reprioritization serves two purposes. First, it allows users to edit the displayed patterns list so that it better reflects their human knowledge: machine-perceived patterns the user knows are irrelevant can be discarded, while important, correctly identified patterns can be "stickied" to the top. A more [adjective here]  list aids the developer in recalling patterns to reapply as they code and conveying their intentions to other developers. Second, reprioritization provides IRIS with feedback on its recognition and application of patterns, resulting in more accurate pattern learning and autocomplete recommendations.

When first learned in training a decision tree, code patterns are initially in the \textit{standard} state. Upvoting a standard pattern transitions it into the \textit{prioritized} state. Prioritized patterns reflect insight that the developer themselves has offered into code patterns. Once in the prioritized state, code patterns are displayed in a separate section of the All Code Patterns panel, enabling developers to see prioritized patterns at a glance (shown in \figurename{ \ref{fig:iris_patterns_priority}}).  
%, and are used to provide autocomplete recommendations. Standard patterns can be promoted to become \textit{prioritized} patterns, which elevates their target features above those of standard patterns in the autocomplete menu. 
Suggestions from prioritized patterns are always listed first in the autocomplete list (\figurename{ \ref{fig:iris_auto_priority}}).
%When the system populates the autocomplete menu, it first checks the prioritized patterns list for applicable recommendations before turning to the decision tree. 

Standard patterns may be demoted to enter the \textit{blacklisted} state. This state reflects the developers' indication that the code pattern does not match their intent. As such, blacklisted code patterns are never shown to the developer as a potential code completion. In addition, to prevent IRIS from learning other similar patterns, any entry in the training data that matches a blacklisted code pattern is removed.
%blacklisted are used to screen training data. Any entry in the  as a means of screening the training set: any relation that matches a blacklisted pattern is removed from the dataset. 
This enables the developer to give feedback on a single code pattern and for IRIS to incorporate this insight more broadly.
%This process prunes the decision tree in a manner reflective of user-provided feedback, generating more desirable code patterns and recommendations.

%(shown in \figurename{\ref{fig:iris_patterns_blacklist}})

Because standard patterns are machine learned, the standard patterns list is continually updated as changes are made to the document. Standard code patterns may be automatically change or removed at any time. 
%Standard patterns may undergo alteration or removal automatically. 
Prioritized or blacklisted code patterns which reflect the developers' intent remain until their priority is edited by the developer.

%Patterns can be promoted or demoted by clicking the up- or down-arrow buttons, respectively, that appear alongside them. Clicking the up-arrow next to a standard pattern promotes it to the prioritized patterns list, presented in a separate table above the standard patterns
%. 
%Moreover, the priority designation is reflected in the autocomplete recommendation order: prioritized patterns' target features appear at the top of the dropdown . 

Developers may also choose to manually enter a prioritized code pattern by clicking the Add button in the All Code Patterns panel. The developer can first choose whether to create a tag, attribute, or attribute value code pattern and then 
%The user may then select a target feature type from the dropdown to patternize. 
create conditions and specify the target.
Custom code patterns must have a target and at least one condition to be valid. 
%As custom code patterns reflect developer intent, they are created as a prioritized pattern.
%After the user has outlined the pattern components as desired, they may add the pattern. Because custom patterns are added directly to the prioritized list (as opposed to the training set of relations), the number of conditions can dramatically alter the scope of the pattern; as more conditions are included, the pattern becomes less generally applicable.
%Clicking the down-arrow next to a standard pattern demotes it to the blacklisted patterns list, which appears below the standard patterns list .

In cases where developers have taken the time to create prioritized or blacklisted code patterns, developers may wish to share these code patterns for use in other HTML documents. Developers may use these rules to capture a look and feel, which might be shared with other developers. 
Developers can click Export to download a JSON file containing prioritized and blacklisted patterns for the current document. Developers may then invoke Import to load code patterns into IRIS.

\section{Evaluation}

\subsection{Method}
To investigate the impact of enabling developers to interact with code patterns, we conducted a user study. We recruited twenty-four participants through personal contacts and social media. All participants had prior experience in web development (average 3.2 years). 17 were male, and 7 female. Participants were paid \$25 for 2 hours of their time. 

Participants were assigned to a control or experimental condition. Participants in the control condition were provided a baseline HTML editor with autocomplete functionality, where code recommendations were generated through the same process as used in IRIS (Section  \ref{subsubsect:autocomplete}). However, the IRIS panel was disabled, and developers were unable to view or edit code patterns. Participants in the experimental condition were provided the same HTML editor augmented with a fully functional IRIS. Participants were each assigned to work on two tasks out of three possible tasks. 

To encompass a range of potential tasks in which developers might benefit from IRIS, we designed three tasks: a \textit{creation}, \textit{continuation}, and \textit{correction} task. This was intended to sample a range of situations in which developers might need to interact with code patterns.
In the \textit{creation} task, participants were directed to build an HTML code document from scratch in accordance with provided specifications. The specifications outlined document elements to create (e.g., \textit{Create a two-column table in the center of the page}) and their expected styling (\textit{Color the header the same as the footer}).
In the \textit{continuation} task, participants were directed to complete an unfinished HTML document in accordance with provided specifications. Participants were given a 400 line HTML document. The specifications asked participants to replicate specific elements (e.g., \textit{Add a third link to the navigation bar}) as well as to restyle the document in specific ways (\textit{Re-style the buttons in the second row to match those in the first}).
In the \textit{correction} task, participants were directed to find and fix inconsistencies in a document. These inconsistencies included missing or incorrect HTML features (e.g. a \texttt{<p>} tag on line 34 should be a \texttt{<caption>}) and relationships (\texttt{div} on line 252 should be nested under a \texttt{aside} parent).

At the beginning of the study, participants were first asked to complete a brief tutorial explaining the main features of the HTML editor and, for experimental participants, IRIS. Participants were then given up to 75 minutes to complete each of the two main tasks.  
Participants were instructed to notify the researcher when they felt they had completed a task. 
%Participants were informed that their work would be evaluated on time taken and success at meeting requirements, 
As participants worked, we collected a screen recording for analysis. 
%and agreed to having their screen recorded for analysis. 
After completing the tasks, we interviewed participants about their experiences.

\subsection{Results}

\begin{table}[tp]
\renewcommand{\arraystretch}{1.4}
\caption{Average task time by task and condition}
\begin{center}
\begin{tabular}{c c c}
&\multicolumn{2}{c}{\textbf{Mean task time}} \\
\cline{2-3}
\textit{Task} & \textit{Control} & \textit{IRIS} \\
\hline
Creation & 40.3 & 34.4 \\
Continuation & 57.1 & 44.1 \\
Correction & 44.1 & 29.4 \\
\end{tabular}
\label{tab1}
\end{center}
\end{table}

\begin{table}[tp]
\renewcommand{\arraystretch}{1.4}
\caption{Average task success by task and condition}
\begin{center}
\begin{tabular}{c c c c}
&\multicolumn{2}{c}{\textbf{Mean success score}}& \\
\cline{2-3}
\textit{Task} & \textit{Control} & \textit{IRIS} & \textit{Max. Possible Score} \\
\hline
Creation & 11.4 & 14.9 & 17 \\
Continuation & 11.4 & 17.0 & 21 \\
Correction & 14.8 & 23.5 & 27 \\
\end{tabular}
\label{tab2}
\end{center}
\end{table}

\subsubsection{Creation Task}
Participants using IRIS finished building the outlined webpage after an average of 40.3 minutes, compared to 34.4 for control participants. However, the Welch's one-tailed \textit{t}-test revealed this difference only approached significance ($p = 0.06$). To evaluate participants' success, we scored each HTML document created, awarding one point for each item of the task specification they successfully completed. Participants with IRIS successfully completed significantly more requirements ($p < 0.01$), completing an average of 14.9 compared to 11.4 by control participants.

Several experimental participants opted to focus on viewing code patterns, rather than editing code patterns. These participants periodically browsed the All Code Patterns list to assess their progress. As one participant explained, \textit{``I didn't really need to highlight samples of code I just wrote. But just having the list there helped me keep track of what work I've already done and what I have left.''} Others noted that seeing the existing patterns in their code was useful for both evaluating task-compliance and conceiving ideas for what to develop next. 
%Several participants with IRIS did not edit code patternn during  throughout the creation task.

Other experimental participants used IRIS to manually define their own code patterns. One participant explained that adding their own patterns ahead of time helped with \textit{``sticking to a plan''}, while another observed that it \textit{``made the autocompletes [sic] more useful''} by transferring his intent to the system. These participants made extensive use of promoting and demoting code patterns, enabling them to remove patterns which overfit the data and focus on intended patterns. 

\subsubsection{Continuation Task}
Participants with IRIS completed the continuation task significantly faster ($p = 0.03$), finishing in 57.1 minutes compared to 44.1 minutes for control participants. 
To score participants' success, we gave participants one point for each specification they satisfied and one or two points for following the specification consistently throughout their document. This led to a maximum score of 21 points. %participants' code for the continuation task. For each specification, one point was awarded on the basis of efficacy (successfully fulfilling the specification) and one to two points for consistency (applying appropriate relationships). 
Experimental participants performed significantly better ($p < 0.01$), with an average progress of 17.0 rather than 11.4.

IRIS enabled participants to more quickly identify and reapply code patterns within their document. Participants with IRIS often relied on autocomplete recommendations to develop code immediately and later, briefly review the Current Code Pattern and highlight usage examples to double check the recommendations' applicability. Compared to control participants with suggested completions but no insight into their source, IRIS participants developed stronger trust in the recommendations and relied on them more heavily over time. In contrast, several control group participants appeared skeptical of the recommendations, either ignoring them or spending substantial time searching for examples to verify them. A few IRIS participants used upvoting or downvoting to edit code patterns. One participant observed that \textit{``The autocomplete got even more accurate as I voted on the patterns''}, making document development \textit{``easier and easier''}. 

%IRIS enabled participants in the continuation task to meet requirements more often and more consistently with existing code. 
IRIS participants made use of the code patterns list to learn code patterns, locating them by highlighting usage examples and reapplying them in new code. Participants used the patterns list to understand the appropriate attribute value to use in various contexts, enabling them to reproduce appropriate attribute-value pairs for a given element. Many participants browsed the code patterns for its parent tag conditions, which aided them in reproducing appropriate parent-child structures. For example, participants first became aware of the rule to nest self-contained \texttt{img} elements inside a \texttt{figure} parent by seeing this code pattern in the list (as a new semantic element introduced in HTML5, the \texttt{<figure>} tag and its usage may not be widely understood). Most participants followed up by highlighting examples of this code pattern, either recreating or copying document snippets involving \texttt{figure} and \texttt{img}. In contrast, control participants often incorrectly created \texttt{img} elements without \texttt{figure} parents, and more generally, were not as aware of the existence of patterns.

\subsubsection{Correction Task}
IRIS participants completed the correction task significantly faster ($p < 0.01$), finishing in an average of 29.4 minutes compared to 44.1 minutes for control participants. To score participants' success, we gave participants one point for the addition, modification, or deletion of code to make it consistent with the code patterns in the document. This led to a maximum score of 27 points. IRIS participants were significantly more successful ($p < 0.01$), with an average score of 23.5 compared to 14.8 for control participants.

A key benefit IRIS offered participants was the ability to highlight pattern examples and violations. All participants with IRIS used the pattern inspector to scan the HTML document for inconsistencies. Participants generally interpreted the red highlights to indicate a defective code feature, and the yellow highlights to suggest a missing or defective attribute-value pair. This heuristic was often helpful for participants in identifying inconsistent code. However, it sometimes misled a few participants into ``fixing'' elements that did not need correction, as certain defects featured predominantly in the document, and thus, were listed by IRIS as code patterns. Despite this occasional shortcoming, the patterns list and inspection tool were instrumental in helping participants successfully find and repair code defects:
\begin{quote}
\textit{``The magnifying glass [button] was very handy. For each pattern I pressed [the button], read the colored lines and compared them... And then figured out which line needed fixing from there.''} (IRIS Participant)
\end{quote}

In contrast, control participants struggled to locate inconsistencies, particularly those concerning HTML values.
\begin{quote}
\textit{``Sorting through all the code was really confusing and time-consuming. I couldn't tell what I was supposed to do for the most part... Only a few blatantly wrong tags stood out to me.''} (Control Participant)
\end{quote}
%In addition, the speed granted by the pattern inspection tool aided the experimental group in completing work faster than the control group.

\section*{Related Work}

Our work builds on prior work in autocomplete tools for developers, systems which identify and make use of patterns in code, and work in explainable AI. 

Autocomplete is one of the most widely used features in modern development environments, with one study finding that 6.7\% of all of developers' IDE commands were code completions \cite{MurphySoftware2006}.
A number of systems have explored techniques for offering more effective autocomplete interactions. 
Many of these systems build a model of document \textit{context} to offer more accurate predictions. 
CSCC offers developers potential method calls as code completions based on the context of surrounding method calls \cite{AsaduzzamanICSME14}.
Dompletion builds a model of an HTML document object model, enabling suggestions of valid completions which respect element types\cite{BajajASE14}. 
Calcite crowdsources the creation of virtual methods, enabling developers to use code completion for methods that developers expect to exist but do not\cite{MootyVLHCC2010}.
Proksch et al. discuss some of the challenges in evaluating method recommender systems \cite{ProkschASE2016}.
Jungloid mining enables developers to find sequences of method calls which convert from an object in one type to an object of a different type\cite{MandelinPLDI2005}. Work has explored using code history information to improve code completion \cite{RobbesASE2008}.
Other tools have explored approaches for completing entire method bodies from prior examples or code clones \cite{HillASE2004}. Active code completion offers the developer palettes for creating specific complex literals\cite{OmarICSE2012}. Beyond code completion, other approaches help connect code more effectively to documentation to help developers more effectively find methods. For example, work has explored using documentation to augment interactions in the IDE\cite{GoldmanVLHCC08}.

A variety of autocomplete systems have built statistical models of documents, using these models to power better autocomplete. These systems largely rely on the inherent high predictability and repetitiveness of code\cite{Hindle2016}.
Systems such as MAPO\cite{Zhong2009}, GraPacc\cite{NguyenICSE2012}, and method call relationship graphs \cite{LiFSE2016} offer approaches for capturing patterns in API method usages.
Other work has explored approaches for learning method completions from code repositories \cite{BruchFSE2009, RaychevPLDI2014}. Recent systems have expanded the amount of interactivity, enabling developers to interactively enter keywords and choose among alternatives to refine completions \cite{RongUIST16}.

Building on the growing fear of AI systems creating a "black box society"\cite{Pasquale2015}, work in the area of explainable AI has explored ways in which difficult to understand machine learning models may be represented and communicated more simply to users\cite{Guidotti2018}. For example, model understanding through space explanations enables translating an arbitrary black-box representation of a machine learning system into decision sets which capture the behavior of the black box in specific circumstances \cite{Lakkaraju2019FaithfulAC}. In interactive machine learning, users can view and correct classifications made by a system\cite{Fails2003}. Techniques such as why-oriented\cite{Kulesza2011} and explanatory debugging\cite{Kulesza2015} enable users to view predictions made by a naive Bayes model and make corrections back to the model. Our work builds on these ideas, applying the idea of interactive machine learning to document editing through code completion and viewing violations as well as exploring the use of rules as a means to view and edit machine learning models. 

\section{Limitations and Threats to Validity}
Our results might vary for larger HTML documents or for documents which have been created over a longer period of time, where the number of code patterns is larger. While the potential value of our approach in managing more patterns may grow, there may also be a larger potential for finding spurious or overlapping patterns. In our study, we observed developers creating a document they began themselves, working to stay consistent, as well as modifying an existing document that they had not written. Our results might vary for developers working in the same document over a period of time who have internalized more code patterns.

%Machine learning traditionally assumes that there is little noise in the data or, if there is noise, the noise is independent of the model. Even in cases where a user is providing the labels based on instructions from the learner on what would be most insightful (active learning), the assumption is that the user is right. But in seeing patterns, as revealed from models and rules learned from the data, the user may change their view on what the correct data is in order to result in a simpler model. This relies on having a setting where the data is generated entirely by the user and there is no other external source of truth on what the correct data is.

% Explaining Collaborative Filtering Recommendations 

% https://www.cc.gatech.edu/~alanwags/DLAI2016/%28Gunning%29%20IJCAI-16%20DLAI%20WS.pdf
% http://openaccess.city.ac.uk/13819/1/paper326.pdf
% https://pdfs.semanticscholar.org/6b84/efd07192cc1ec87c32b93a335449a365d045.pdf
% https://ai.intel.com/the-challenges-and-opportunities-of-explainable-ai/
% https://www.nytimes.com/2017/11/21/magazine/can-ai-be-taught-to-explain-itself.html
% https://distill.pub/2017/feature-visualization/
% http://pages.cs.wisc.edu/~anhai/papers1/hilda18.pdf

% Interactive machine learning systems first use ML, then use crowd to edit results:
% https://dl.acm.org/citation.cfm?id=3158226
% Special issue on interactive ML: https://dl.acm.org/citation.cfm?id=3232718&picked=prox
% Paper introducing idea of interactive ML: https://dl.acm.org/citation.cfm?id=604056   http://byu.danrolsenjr.org/paperPDFs/InteractiveMachineLearning.pdf

\section{Discussion}
In this paper, we proposed an approach for \textit{editable} AI, in which the human and computer work together to create and maintain code patterns. Code patterns are first learned by the computer through decision trees. Code patterns are then used to power code completions, suggesting ways in which developers may write code which follows code patterns. To help developers explain code patterns, developers can view a compact representation of code patterns, comparing alternative predictions and viewing examples to see the source of predictions. Developers may use patterns to identify violations, viewing examples in the document which violate patterns. In cases where learned code patterns overfit the document and do not reflect developers' true intent, developers may downvote code patterns, which the system can then use to prune training data to prevent similar rules from being learned. Or developers may signal that a code pattern matches their intent by upvoting or authoring their own code patterns.

Our results reveal that editablity and explainability matter. Compared to developers offered autocomplete with the same learned patterns, developers with the ability to view and edit code patterns through IRIS were able to edit and correct HTML documents more quickly and create, edit, and correct HTML documents more successfully. Developers were able to use code patterns to explain suggested completions, gathering examples to see the source of these recommendations and increasing their trust in the system. Developers used code patterns to identify and correct violations. As developers developed their own intent, developers reflected this by editing code patterns, enabling autocomplete to offer better recommendations. In this way, editability and explainability helped developers to author more consistent and well-structured documents. This suggests the potential benefit, for creative work, of offering the user greater control over machine learning. 

%We chose the continuation and correction tasks to assess key motivations behind IRIS, including surfacing existing patterns to help developers re-apply them in future work, and highlighting pattern examples and violations to help developers resolve defects. The creation task was intended to evaluate whether the benefits of IRIS depend on a well-developed codebase, or can be realized as a codebase is being developed. 

Code patterns offer a way of capturing the look and feel of a document. By representing the structure of the elements which exist in an HTML document, developers can more directly understand and interact with the visual look and feel. For example, code patterns might capture that \texttt{img} elements in the header should each have a particular size as expressed by a \texttt{class}. Our representation of code patterns is limited in a number of respects. In representing code patterns, we included features such as parent tags and attributes to capture the context-specific nature of HTML document structure. But additional contextual features likely sometimes matter, from the numeric order of an element to the characteristics of other ancestor elements. More expressive representations might capture these. Code patterns are complimentary to CSS, helping capture document structure rather than only visual styling. Compared to CSS styles, code patterns are more flexible, as unlike CSS styles which automatically apply to elements, developers are free to create elements whether or not they follow a code pattern. However, there may be cases where code patterns and CSS styles more directly overlap, and it may sometimes be helpful to enable code patterns to be converted to CSS rules or vice versa. 

Our approach might also be applied to other documents where the user has intent in mind about its structure. Most directly, our approach might be applied to code written in a programming language. However, such patterns may be considerably more complex, involving a much wider variety of condition features and targets. While existing statistical approaches to modeling code may offer hints into appropriate representations, more work remains to understand appropriate ways of learning these patterns and how they might be succinctly communicated to developers.

%making it more challenging to identify 

%How this approach might be applied to code rules more generally. How this is more challenging]

%Could envision applying this approach to code. Patterns space is far more complex, requiring careful consideration of what features would be used. 

%The problem we were solving is x...
%We found that IRIS helped developers to do a, b, c...

%\textbf{Todo:\\
%Reasons why IRIS didn't result in significant reduction in Creation task duration:\\
%1. Autocomplete kicked in less, since there's less training data to draw from (possible solution: train on more than just the current document being developed?)
%\\
%2. Many of the Creation participants' use of IRIS consisted primarily of checking the All Patterns list agains the requirements to look for discrepancies - Control group didn't have this technique at their disposal, which cost them points but also didn't eat up some time like it did for experimental}

%\section{Conclusion}

\section*{Acknowledgments}
We thank our study participants for their time. This work was conducted in part while Kartik Chugh was an intern in the Aspiring Scientists Summer Internship Program at George Mason University. Andrea Solis was supported in part by an Undergraduate Research in Educational Data Mining grant, NSF IIS-1757064.

\bibliographystyle{IEEEtran}
\bibliography{latoza}

\end{document}